\documentclass[a4paper,11pt]{article}

\usepackage{amsmath,amsthm,amsfonts,amssymb, graphicx,mathabx,epsfig}
\usepackage{bbm}
\usepackage{authblk}
\usepackage{bm}
\usepackage{graphicx,color}
\usepackage{footmisc}
\usepackage{epstopdf}
\usepackage{dsfont}
\usepackage{pgfplots}
\usepackage{subfig}
\usetikzlibrary{fit}
\usepackage{slashed}

\usepackage{graphicx}
\usepackage{dcolumn}
\usepackage{bm}
\usepackage{epstopdf}

\newcount\driver
\newcount\bozza

scaled\magstep1                  
scaled\magstep1 \font\msytw=msbm10 scaled\magstep1
 
scaled\magstep1                 \font\indbf=cmbx10 scaled\magstep2

scaled \magstep2

{\count255=\time\divide\count255 by 60
\xdef\hourmin{\number\count255}
        \multiply\count255 by-60\advance\count255 by\time
   \xdef\hourmin{\hourmin:\ifnum\count255<10 0\fi\the\count255}}

\let\a=\alpha \let\b=\beta    \let\g=\gamma     \let\d=\delta     \let\e=\varepsilon
\let\z=\zeta  \let\h=\eta     \let\th=\vartheta \let\k=\kappa     \let\l=\lambda
\let\m=\mu    \let\n=\nu              \let\p=\pi        \let\r=\rho
\let\s=\sigma \let\t=\tau            
\let\ps=\psi   \let\o=\omega     
\let\G=\Gamma \let\D=\Delta       \let\L=\Lambda

\def\ZZ{{\cal Z}}

\def\pp{{\bf p}}\def\qq{{\bf q}}\def\xx{{\bf x}}
  
\def\yy{{\bf y}}\def\kk{{\bf k}}\def\nn{{\bf n}}

       \def\oo{{\underline \omega}}
\def\ee{{\underline \varepsilon}}

        \def\ZZZ{\hbox{\msytw Z}}

\let\==\equiv

\let\io=\infty
\let\0=\noindent

\def\*{{\hfill\break\null\hfill\break}}

\def\media#1{{\langle#1\rangle}}

\def\tilde#1{{\widetilde #1}}

\def\tende#1{\,\vtop{\ialign{##\crcr\rightarrowfill\crcr
             \noalign{\kern-1pt\nointerlineskip}
             \hskip3.pt${\scriptstyle #1}$\hskip3.pt\crcr}}\,}
\def\otto{\,{\kern-1.truept\leftarrow\kern-5.truept\to\kern-1.truept}\,}

\def\wh#1{\widehat{#1}}
\def\hat#1{\wh{#1}}
\def\sqt[#1]#2{\root #1\of {#2}}

\def\bp{{\bar \ps}}

\def\ZZ{{\cal Z}}

\def\T#1{{#1_{\kern-3pt\lower7pt\hbox{$\widetilde{}$}}\kern3pt}}
\def\VVV#1{{\underline #1}_{\kern-3pt
\lower7pt\hbox{$\widetilde{}$}}\kern3pt\,}
\def\W#1{#1_{\kern-3pt\lower7.5pt\hbox{$\widetilde{}$}}\kern2pt\,}

\def\indica{\leaders \hbox to 0.5cm{\hss.\hss}\hfill}
\def\guida{\leaders\hbox to 1em{\hss.\hss}\hfill}
\mathchardef\oo= "0521

\def\V#1{{\bf #1}}
\def\pp{{\bf p}}\def\qq{{\bf q}}\def\xx{{\bf x}}
\def\yy{{\bf y}}\def\kk{{\bf k}}\def\nn{{\bf n}}

\def\oo{{\underline \omega}}

\def\qed{\raise1pt\hbox{\vrule height5pt width5pt depth0pt}}
 
  \def\bp{{\bar p}} 

\def\indic{\hbox{\raise-2pt \hbox{\indbf 1}}}

 \def\ZZZ{\hbox{\msytw Z}}

\def\ins#1#2#3{\vbox to0pt{\kern-#2 \hbox{\kern#1 #3}\vss}\nointerlineskip}

\newdimen\xshift \newdimen\xwidth \newdimen\yshift

\def\insertplot#1#2#3#4#5#6{%
\xwidth=#1pt \xshift=\hsize \advance\xshift by-\xwidth \divide\xshift by 2%
\begin{figure}[ht]
\vspace{#2pt} \hspace{\xshift}
\begin{minipage}{#1pt}
#3 \ifnum\driver=1 \griglia=#6
\ifnum\griglia=1 \openout13=griglia.ps \write13{gsave .2
setlinewidth} \write13{0 10 #1 {dup 0 moveto #2 lineto } for}
\write13{0 10 #2 {dup 0 exch moveto #1 exch lineto } for}
\write13{stroke} \write13{.5 setlinewidth} \write13{0 50 #1 {dup 0
moveto #2 lineto } for} \write13{0 50 #2 {dup 0 exch moveto #1
exch lineto } for} \write13{stroke grestore} \closeout13
\includegraphics{griglia.ps} \fi
\includegraphics{#4.ps}\fi%
\ifnum\driver=2 \fi
\end{minipage}
\caption{#5}
\end{figure}
}

\newdimen\shift \shift=-1.5truecm
\def\lb#1{%
\ifnum\bozza=1
\label{#1}\rlap{\hbox{\hskip\shift$\scriptstyle#1$}}
\else\label{#1} \fi}

\def\be{\begin{equation}}
\def\ee{\end{equation}}
\def\bea{\begin{eqnarray}}\def\eea{\end{eqnarray}}
\def\bean{\begin{eqnarray*}}\def\eean{\end{eqnarray*}}
\def\bfr{\begin{flushright}}\def\efr{\end{flushright}}
\def\bc{\begin{center}}\def\ec{\end{center}}
\def\bal{\begin{align}}\def\eal{\end{align}}
\def\ba#1{\begin{array}{#1}} \def\ea{\end{array}}
\def\bd{\begin{description}}\def\ed{\end{description}}
\def\bv{\begin{verbatim}}\def\ev{\end{verbatim}}
\def\nn{\nonumber}
\def\Halmos{\hfill\vrule height10pt width4pt depth2pt \par\hbox to \hsize{}}
\def\pref#1{(\ref{#1})}

\def\ins#1#2#3{\vbox to0pt{\kern-#2 \hbox{\kern#1 #3}\vss}\nointerlineskip}

\newdimen\xshift \newdimen\xwidth \newdimen\yshift
\newcount\griglia

\def\insertplot#1#2#3#4#5#6{%
\xwidth=#1pt \xshift=\hsize \advance\xshift by-\xwidth \divide\xshift by 2%
\begin{figure}[ht]
\vspace{#2pt} \hspace{\xshift}
\begin{minipage}{#1pt}
#3 \ifnum\driver=1 \griglia=#6
\ifnum\griglia=1 \openout13=griglia.ps \write13{gsave .2
setlinewidth} \write13{0 10 #1 {dup 0 moveto #2 lineto } for}
\write13{0 10 #2 {dup 0 exch moveto #1 exch lineto } for}
\write13{stroke} \write13{.5 setlinewidth} \write13{0 50 #1 {dup 0
moveto #2 lineto } for} \write13{0 50 #2 {dup 0 exch moveto #1
exch lineto } for} \write13{stroke grestore} \closeout13
\includegraphics{griglia.ps} \fi
\includegraphics{#4.ps}\fi%
\ifnum\driver=2 \fi
\end{minipage}
\caption{#5}
\end{figure}
}

\newdimen\shift \shift=-1.5truecm
\def\lb#1{%
\label{#1}\rlap{\hbox{\hskip\shift$\scriptstyle#1$}}
\else\label{#1} \fi}

\def\be{\begin{equation}}
\def\ee{\end{equation}}
\def\bea{\begin{eqnarray}}\def\eea{\end{eqnarray}}
\def\bean{\begin{eqnarray*}}\def\eean{\end{eqnarray*}}
\def\bfr{\begin{flushright}}\def\efr{\end{flushright}}
\def\bc{\begin{center}}\def\ec{\end{center}}
\def\bal{\begin{align}}\def\eal{\end{align}}
\def\ba#1{\begin{array}{#1}} \def\ea{\end{array}}
\def\bd{\begin{description}}\def\ed{\end{description}}
\def\bv{\begin{verbatim}}\def\ev{\end{verbatim}}
\def\nn{\nonumber}
\def\Halmos{\hfill\vrule height10pt width4pt depth2pt \par\hbox to \hsize{}}
\def\pref#1{(\ref{#1})}

\usepackage{amsmath}
\usepackage{amsfonts}
\usepackage{amssymb}
\usepackage{epstopdf}

\font\msytw=msbm9 scaled\magstep1 
scaled\magstep1

\let\a=\alpha \let\b=\beta  \let\g=\gamma  \let\d=\delta
\let\e=\varepsilon
\let\z=\zeta  \let\h=\eta   \let\th=\theta \let\k=\kappa \let\l=\lambda
\let\m=\mu    \let\n=\nu         \let\p=\pi    \let\r=\rho
\let\s=\sigma \let\t=\tau    
\let\ps=\Psi   \let\o=\omega
\let\G=\Gamma \let\D=\Delta  \let\L=\Lambda

\def\qq{{\bf q}} \def\pp{{\bf p}}
 \def\xx{{\bf x}} \def\yy{{\bf y}} 
\def\kk{{\bf k}}

\def\nn{\nonumber}

 \def\ZZZ{\hbox{\msytw Z}}

\def\\{\hfill\break}
\def\={:=}
\let\io=\infty
\let\0=\noindent
\def\media#1{{\langle#1\rangle}}

\def\tende#1{\,\vtop{\ialign{##\crcr\rightarrowfill\crcr\noalign{\kern-1pt
    \nointerlineskip} \hskip3.pt${\scriptstyle #1}$\hskip3.pt\crcr}}\,}
\def\otto{\,{\kern-1.truept\leftarrow\kern-5.truept\to\kern-1.truept}\,}

\def\wh{\widehat}
\def\to{\rightarrow}

\def\qed{\hfill\raise1pt\hbox{\vrule height5pt width5pt depth0pt}}

\def\V#1{{\bf#1}}
\def\be{\begin{equation}}
\def\ee{\end{equation}}
\def\bp{\begin{pmatrix}}
\def\ep{\end{pmatrix}}
\def\bea{\begin{eqnarray}}
\def\eea{\end{eqnarray}}
\def\nn{\nonumber}
\def\pref#1{(\ref{#1})}

\def\lb{\label}

\begin{document}

\title{Emergent  Adler-Bardeen theorem}

\author[1]{Vieri Mastropietro}
\affil[1]{University of Milano, Department of Mathematics ``F. Enriquez'', Via C. Saldini 50, 20133 Milano, Italy}

\maketitle

\begin{abstract} 
We consider a QED$_{d+1}$, $d=1,3$ lattice model 
with emergent
Lorentz or chiral symmetry, both when the interaction is irrelevant or marginal. 
While the correlations present symmetry breaking corrections, we prove
that the Adler-Bardeen (AB) non-renormalization property holds at a non-perturbative level even at finite lattice: all radiative corrections to the anomaly are vanishing. The analysis uses a new technique
based on the combination of non-perturbative regularity properties
obtained by exact renormalization Group methods and Ward Identities. 
The AB property, essential for the renormalizability of the standard model, is therefore a robust
feature imposing no constraints on possible symmetry breaking terms, at least in the class of lattice models considered.
\end{abstract}
\maketitle

\maketitle

\section{Introduction}
According to modern understanding 
several symmetries of 
particle physics can be approximate and emergent, see e.g. \cite{Pol}, \cite{Po},\cite{Wi},
and possibly broken at the Planck length scale. The Adler-Bardeen (AB) non-renormalization property \cite{A},\cite{AB},\cite{AB1} 
is essential to ensure the renormalizability of the Standard model, through the anomaly cancellation.
The proof of the AB property is based on identities between class of graphs and relies on
the validity of Lorentz and chiral symmetry. If the symmetry breaking terms are dimensionally irrelevant, one expects 
that the corrections are of the order 
of the Planck divided by the particle physics length scale, hence typically undetectable.
However this would be not true if corrections are present to the
AB property; even very small radiative contributions  
would be amplified if they break renormalizability.
It is therefore interesting to see if the anomaly non-renormalization holds generically even
when symmetry breaking terms are present at the Planck length scale, or if in contrast its validity requires that they are absent or at least of special form.

We consider the simple situation where the symmetry violation is produced by a lattice, with
spacing small compared to particle physics
lengths but large or comparable to the Planck length scale; lattice models are often used to mimic the violation of symmetries, see 
e.g.\cite{P0}, \cite{P1},\cite{P2},\cite{P3}. In particular, the model we consider
is the interacting extension of
the Nielsen-Ninomiya simulation of the chiral anomaly \cite{NN},
that is lattice fermions coupled with a quantum massive photon field, with
an emerging description in terms of massless QED$_{d+1}$, $d=1,3$. There are 
corrections to the Lorentz invariant part of the correlations which are {\it 
non-vanishing} and of order of the momentum times the lattice spacing.
In contrast, we prove that
the anomaly is perfectly {\it
non-renormalized }, even in presence of finite symmetry breaking terms; that is, at least in the class of lattice models we consider,
{\it the AB non-renormalization is a robust feature imposing no constraints on the symmetry breaking terms}.

Our results are based on a novel technique 
based on the combination of constructive regularity properties
obtained by exact Renormalization Group (RG) methods 
and Ward Identities. 
The contribution of irrelevant terms at each step is essential and fully taken 
into account. The results are fully non-perturbative, as
physical quantities are expressed in terms of series
whose {\it convergence} is established in presence of a finite photon mass, as 
consequence of cancellations due to Pauli principle
(see \cite{M1} for an introduction to such methods).
This is a major difference with respect to other approaches to the anomaly which give results valid only order by order, see e.g. 
\cite{K},\cite{K1}.
The strategy of proof was used in 
\cite{GMP} for {\it irrelevant} interactions and is here extended to the {\it marginal} case. 
Even if the validity of the AB property is proved for the $U(1)$ gauge group and with a photon mass regularization, the 
result indicates that the anomaly cancellation condition is common between the continuum theory and the lattice theory and 
suggests that anomaly-free chiral gauge theory, as the Standard model, can be formulated non-perturbatively by lattice formulation.

\section{Lattice models and anomaly non-renormalization}

The massless lattice QED$_{d+1}$ model we consider is the interacting extension of the Nielsen-Ninomiya anomaly simulation
 \cite{NN}, where the interaction with a quantum photon field is included.
The detailed form of the lattice has no importance and we do a specific choice just for definiteness.

In $d=1$ we consider 
a linear lattice $\L=\{x=n a,  n\in \ZZZ\})$.
If $\psi^\pm_x$, $x\in\L$,  are fermionic creation or annihilation operators defined on the Fock space and verifying $\{\psi^+_x,\psi^-_y\}=\d_{x,y}$, 
$\{\psi^+_x,\psi^+_y\}=\{\psi^-_x,\psi^-_y\}=0$,
the  lattice
Hamiltonian is \be H_0={t\over a}\sum_{x}
\Big({1\over 2}(\psi^+_{x+a} \psi^-_x+\psi^+_x \psi^-_{x+a})-\cos(\z ) \psi^+_x \psi^-_x\Big)\label{h11} \ee
We define
\be
\hat\psi^\pm_k=\sum_{x\in \L} e^{\mp i k x} \psi^\pm_x\quad\quad  
\psi^\pm_x=\int {d k\over (2\pi)} e^{\pm i k x} 
\hat\psi^\pm_k
\ee
where $|k|\le \pi/a$
and
\be
H_0=\int {d k\over (2\pi)} h(k) \hat\psi^+_k \hat\psi^-_k \quad\quad h(k)={t\over a}(\cos k a- \cos \z).
\ee
Note that for $k=\pm \z/a+q$ one has $h(k)= \pm v q+O(q^2 a)$, $v=t\sin \z$, that is the dispersion relation is essentially linear.

In $d=3$ we consider a lattice formed by two sublattices: one is
$\L_1=\{ n_1 a (\frac1{\sqrt2},\frac1{\sqrt2},0)+n_2 a(-\frac1{\sqrt2},\frac1{\sqrt2},0)+n_3 a (0,0,1),\ n_i\in\mathbb Z\}$, and the other is $\L_2=\L_1+a \frac1{\sqrt2}e_1\equiv \L_1+a \d_1$ (we let $e_1,e_2,e_3$ be the elements of the standard Euclidean basis
and $\d_1=\frac1{\sqrt2}e_1$, $\d_2=\frac1{\sqrt2}e_2$).
We associate a fermionic field to each sublattice $\psi^\pm_{x,j}$ with $x\in\L_j$, $j=1,2$ and we consider the Hamiltonian
\bea 
&&H_0=-\frac{t}{2a}\sum_{x\in\L_1}
\Big[ \big(i \psi^+_{x,1} \psi^-_{x+a \d_1 ,2}+i\psi^+_{x-a \d_1,2}
\psi^-_{x,1}+c.c.\big)+\big(\psi^+_{x,1} \psi^-_{x+a\d_2  ,2}-\psi^+_{x-a\d_2 ,2}\psi^-_{x,1}+c.c.\big)\Big]-\nn\\
&&t'
\sum_{x\in\L_1}\Big[ \frac{1}{2a}(\psi^+_{x,1} \psi^-_{x+a e_3 ,1}+ \psi^+_{x,1} 
 \psi^-_{x-a e_3,1})-(\cos (\z)-\frac{1}{a}) 
\psi^+_{x,1}  \psi^-_{x,1})\Big]+\nn\\
&&t'\sum_{x\in\L_2}\Big[ 
\frac{1}{2a}(\psi^+_{x,2} \psi^-_{x+a e_3,2}+\psi^+_{x,2} \psi^-_{x-a e_3,2})-
(\cos (\z)-\frac{1}{a}) \psi^+_{x,2} \psi^-_{x,2})\Big]+\nn\\
&&\frac{t'}{4 a}\sum_{x\in\L_1}\psi^+_{x,1}\Big(\psi^-_{x+a\d_1+a\d_2,1}+
\psi^-_{x+a \d_1-a \d_2,1}+\psi^-_{x-a \d_1+a \d_2,1}+\psi^-_{x-a \d_1-a \d_2,1}\Big)-\label{h12}\\ && \frac{t'}{4 a}
\sum_{x\in\L_2}\psi^+_{x,2}\Big(\psi^-_{x+a \d_1+a\d_2,2}+\psi^-_{x+a\d_1-a\d_2,2}+\psi^-_{x-a \d_1+a\d_2,2}+\psi^-_{x-a\d_1-a\d_2,2}\Big)\nn\eea
We pass to Fourier space :
\be \psi^\pm_{x,1}=\int \frac{dk}{(2\p)^3}e^{\pm ikx}\hat \psi^\pm_{k,1}\qquad \psi^\pm_{x,2}=\int \frac{dk}{(2\p)^3}e^{\pm ikx}\hat \psi^\pm_{k,2}\ee
where the integrals are over the first Brillouin zone 
so that the Hamiltonian reads:
\be H_0=\int \frac{dk}{(2\p)^3}\hat \psi^+_k\begin{pmatrix} \a(k) & \b(k)\\ \b^*(k)& -\a(k)\end{pmatrix}\hat \psi^-_k\equiv 
\int \frac{dk}{(2\p)^3}\hat \psi^+_k h(k) \hat \psi^-_k, \ee
where
\bea && \a(k)= -{t'\over a}(\cos k_3 a-\cos \z) +{t'\over a}(\cos {\scriptstyle\frac{ a k_1}{\sqrt2}}\cos {\scriptstyle\frac{a k_2}{\sqrt2}}-1),\\ 
&& \b(k)=-{t\over a}\sin {\scriptstyle\frac{k_1}{\sqrt2}}+i {t\over a}\sin {\scriptstyle\frac{k_2}{\sqrt2}}.
\eea
If $k_1=q_1$, $k_2=q_2$ and $k_3=\pm \z/a +q_3$  we can write
$h(k)= -v_1 \s_1  q_1-v_1 \s_2  q_2 \mp  v \s_3 q_3+O(a|q|^2)$
with $v_1=t/\sqrt{2}$,
 $v=t' \sin \z$ and $q=(q_1,q_2,q_3)$.
We denote by $\hat\psi_{x_0,k,i}= e^{H_0 x_0} 
\hat\psi^\pm_{k,i} e^{-H_0 x_0}$ where $x_0$ is the Euclidean time;
we introduce the 2-point function $\hat g_{i,j}(\kk)\equiv <\hat\psi^+_{\kk,i}\hat\psi_{\kk,j}^-
>$ given by the following matrix 
\be\hat g(\kk)={1\over 
-i k_0 I+h(k)}\label{prop} \ee
where $\kk=(k_0,k)$. A similar expression holds in $d=1$. The Fourier transform of the propagator $\hat g(\kk)$ is denoted by $g(\xx)$ with $\xx=(x_0,x)$.

It is well known that the above lattice models admit an emerging description in terms 
of Dirac particles \cite{NN}. Indeed 
a Dirac massless particle has 
propagator $<\bar\Psi_\kk\Psi_\kk>={1\over- i \not\kk}$, where
$\slashed{\kk}=\g_\m k_\m$, $\m=0,1,..d$ and
with  $\{\g_\m,\g_\n\} =2 \d_{\m,\n}$ and
the Euclidean Dirac derivative is $\slashed{\partial}$. In $d=1$ a possible realization of $\g$-matrices is 
$\s_1=\g_0$, $\s_2=\g_1$ and
$\s_3=\g_5$ with $\s_1=\begin{pmatrix}&0&1\\
          &1&0\end{pmatrix}
 \quad
\s_2=\begin{pmatrix}
&0&-i\\ &i&0
\end{pmatrix}
\quad\s_3=\begin{pmatrix}&1&0\\
          &0&-1\end{pmatrix}$.
Similarly in $d=3$ 
$$\g_0= \begin{pmatrix} 0 & I \\ I &0 \end{pmatrix}\quad \g_j= \begin{pmatrix} 0 & i\s_j \\-i\s_j &0 \end{pmatrix}, \quad\g_5=\begin{pmatrix}&I&0\\
          &0&-I\end{pmatrix}$$ 
In addition to Lorentz invariance, Dirac particles verify gauge and chiral symmetry,
implying the conservations of the
$d+1$ current  $J_\m=\bar\Psi\g_\m\Psi$ and axial current
$J_\m=\bar\Psi\g_\m\g_5\Psi$. It is also convenient to write
$\Psi=(\Psi_+,\Psi_-)$ and $\bar\Psi=\Psi^+\g_0$, so that the Dirac propagator can be written as, $\hat g^D_{\pm}(\kk)=<\Psi^+_{\pm,\kk} \Psi_{\pm,\kk}>$
\be
\hat g^D_{\pm}(\kk)=(-i k_0\pm k)^{-1}\quad\quad\quad \hat g^D_{\pm}(\kk)=( -i k_0\mp \s_1  k_1\mp\s_2  k_2 \mp \s_3 k_3
)^{-1}
\ee
in $d=1$ and $d=3$ respectively. 
 
Let us look at the lattice propagator \pref{prop} 
restricting to momenta close to $\pm \z/a$. 
We introduce a smooth compact support function 
$\chi_\o(\kk)$, with $\o=\pm$ non vanishing only for $|\kk-\o \bar\z/a |\le 1/(10 a)$ with 
$\bar\z=(0,\z)$ in $d=1$ or $\bar\z=(0,0,0,\z)$ in $d=3$.
In $d=1$ we define
$\hat g_\o(\kk-\o \bar\z/a)=\chi_\o(\kk) \hat g(\kk)$ and
in $d=3$ we
define 
\be 
\hat g_{i,j;\o}(\kk- \o\bar\z/a)=\d_{i,j,\o} 
\chi_\o(\kk) \hat g_{i,j} (\kk)\ee 
with $\d_{i,j,+}=1$,  $\d_{i,j,-}=(-1)^{i+j}$.
The function $\hat g_\o$ is the propagator restricted to momenta around $\pm\bar\z/a$ and,
setting $v=v_1=1$ we get, calling $\kk-\o \bar\z/a=\qq$
\be
\hat g_\pm(\qq)=\hat g^D_{\pm}(\qq)(1+r_\pm(\qq))\quad\quad |r_\pm(\qq)|\le C a |\qq|\label{ro}
\ee
The lattice models \pref{h11} and \pref{h12} admit therefore an emerging description in terms of massless Dirac partcles;
the propagator for momenta far from the inverse spacing
has a Lorentz invariant part up to corrections which are small but non vanishing.
Let us see what happens to the conservation of the currents.
The current in a lattice theory can be introduced using 
the {\it Peierls substitution}. In $d=1$ one introduces an interaction with an external gauge field by writing
\be H_0(A)={t\over a}\sum_{x}
({1\over 2}(\psi^+_{x+a}  e^{i \int_{x+a}^{x} ds A(s)   }
\psi^-_x+\psi^+_x e^{i \int_{x}^{x+a} ds A(s)   }
\psi^-_{x+a})-\cos(\z ) \psi^+_x \psi^-_x)\label{h11a} \ee 
with a similar expression holding in $d=3$;  
the current is defined as
$j_x={\partial H_0(A)\over \partial A(x)}|_{A(x)=0}$,
and
the lattice density is $\r_x=\psi^+_x\psi^-_x$, and they can be combined in $j_\m=(\r,j_1,..,j_d)$.
The lattice density and current vertex are close, in the sense of correlations and
up to corrections as in \pref{ro}, to the Dirac ones
$\bar\Psi\g_\m\Psi$. Such corrections however
do not prevent the conservation of the lattice current
in the sense of Ward Identities (see
\pref{ss} below), as the Peierls substitution ensures gauge invariance at a lattice level.
%

A different situation is encountered in the case of chiral currents.
Following \cite{NN} one can indeed define an analogue of the chiral density and current in the lattice model, by the requirement
that it is close to the Dirac chiral current $\bar\Psi\g_\m\g_5\Psi$
in the sense of correlations, up to corrections. The lattice chiral density can be defined 
as the difference of densities of fermions around $\pm{\bar \z}/a$, that is in $d=1$
\be
\hat\r^{5}_p=\int {dk \over (2\pi)} {\sin ka\over \sin\z} \hat\psi^+_{k+p} \hat\psi^-_k\ee
or  $\hat\r^{5}_p=\int  {dk \over (2\pi)^3} {\sin k_3 a\over \sin\z}\hat\psi^+_{k+p} \hat\psi^-_k$; in coordinate space $\r^{5}_x=-{i  \over 2 \sin \z}
(\psi^+_x \psi^-_{x+a e_3}-\psi^+_x \psi^-_{x-a e_3})$ or $\r^{5}_x=-{i  \over 2\sin\z }
(\psi^+_x \psi^-_{x+a }-\psi^+_x \psi^-_{x-a})$.
The definition of the axial current is given in a similar way inserting a factor
$\sin k a$  or
$\sin k_3 a$ in the Fourier transform of the current.
The axial symmetry is however broken and there is no conservation of axial current.

Let us introduce now a dynamical photon field $\bar A_\m(\xx)$ (not to be confused with the external field 
$A_\m$) with integration $P(d\bar A)$ and propagator \be
v_{\m,\n}(\xx)=\d_{\m,\n} v(\xx)=
\int{d\kk\over (2\pi)^{d+1}  } \chi(\kk){e^{i\kk \xx}  \over \kk^2+M_a^2} 
\d_{\m,\n}
\ee
where $\xx=(x_0,x)$,
$\chi(\kk)$ is a cut-off function vanishing for momenta larger than $O(1/a)$ and $M_a=M$ in $d=1$ and $M_a=a^{-1} M$ in $d=3$ is a regularizing mass (such a regularization is the one adopted in \cite{AB}). For a non-perturbative analysis we find convenient to integrate out the boson field getting a purely fermionic theory, 
that is \be \int P(d\bar A) e^{e\int d\xx \bar A_\m j_\m} =e^{e^2\int d\xx d\yy v(\xx-\yy) j_\m(\xx)  j_\m(\yy)}\ee

The lattice model we consider is therefore defined by the following generating function
\be
e^{W(A_\m,A^5_\m,\phi)}=\int P(d\psi) e^ {V(\psi,A_\m,A^5_\m,\phi)}\label{gen} 
\ee
where $\psi^\pm_{\xx,i}$ (in $d=3$ $i=1,2$ while in $d=1$ $i=1$ and $\psi^\pm_{\xx,1}=\psi^\pm_{\xx}$)
is a set of Grassmann variables $\{\psi^\e_{\xx,i},\psi^{\e'}_{\yy,j}\}=0$, $\e,\e'=\pm$ (with abuse of
notation we denote the Grassmann variables with the same symbol as fields),
$P(d\psi)$ is the fermionic integration with propagator \pref{prop} and
\be
V(\psi, A_\m,A^5_\m,\phi)=
\l\int d\xx d\yy v_{\m,\n}(\xx,\yy) j_{\m,\xx}(A) j_{\n,\yy}(A)+\n N+
B(\psi,A)+\int d\xx A^5_{\m,\xx} j^5_{\m,\xx}(A) 
\ee
where the first term is the interaction, $j_\m=(\r,j_1,..,j_d)$ are the lattice density and current 
expressed in terms of Grassmann variables, $j_\m(A)$ is obtained by $j_\m$ by the Perierls substitution, 
$\int d\xx$ is a notation for $\int dx_0 \sum_x$,
$\l=e^2$ is the coupling and the second term is a counterterm to fix the singularity of the propagator,
$N=\int d\xx \psi^+_\xx \psi^-_\xx$  in $d=1$ or  $N=\int d\xx (\psi^+_{1,\xx} \psi^-_{1,\xx}-\psi^+_{2,\xx} \psi^-_{2,\xx})$ 
in $d=3$. Finally $A_\m,A_\m^5,\phi$ are external fields ($\phi$ is a Grassman variable)
and derivatives of $W$ with respect to $A_\m, A_\m^5,\phi$
give the correlations of the current, chiral current or fermionic field respectively. In order 
to ensure gauge invariance for the external field $A_\m$ (see \pref{xx} below) we define
\be
B(\psi,A)=\int d\xx A_{0,\xx}\r_\xx -(H_0(A)-H_0(0))+\sum_{\e=\pm}\int d\xx \psi^{\e}_\xx\phi^{-\e}_\xx
\ee
with $H_0(A)$ given by \pref{h11a} with Grassmann variables replacing fields
and $j^5_{\m,\xx}(A)$ is obtained by  $j^5_{\m,\xx}$
by the Peierls substitution; in particular the gauge invariant chiral density is 
\be
\r^{5}_\xx(A)=\ZZ^5_0{1\over \sin\z}
(\psi^+_\xx e^{i  \int_{x_3}^{x_3+a } ds A_3(s) }
 \psi^-_{\xx+a e_3}-\psi^+_{\xx-a e_3} e^{i  \int^{x_3}_{x_3-a  } ds A_3(s)   }
\psi^-_{\xx})
\ee
with $A_3(s)=A_3(x_0,x_1,x_2,s)$ and $\ZZ^5_0$ is a renormalization to be properly fixed, see below;
a similar expression holds for the axial current $j^5_{i,\xx}(A)$, and
$\ZZ^5_i$ are the corresponding renormalizations.

The  correlations are obtained by differentiating the generating function with respect to the external fields; in particular
\be
\hat G_2(\kk)={\partial^2 W \over \partial\hat\phi^-_\kk\partial\hat\phi^+_\kk}|_0\quad\quad 
\hat G_{2,1}(\pp, \kk)={\partial^3 W \over \partial \hat A_{\m,\pp} \partial \hat\phi^-_\kk\partial\hat\phi^+_{\kk+\pp}}|_0
\quad\quad  \hat G_{2,1}^5(\pp, \kk)={\partial^3 W \over \partial \hat A^5_{\m,\pp} \partial\hat\phi^-_\kk\partial\hat\phi^+_{\kk+\pp}}|_0
\ee
where given a function $f(A_\m,A^5_\m,\phi)$ we denote $f(A_\m,A^5_\m,\phi)|_0=f(0,0,0)$ and $\hat A_\m$ is the Fourier transform of $A_\m$.

We define 
in $d=1$ $\hat G_\o(\kk-\o \bar\z/a)=\chi_\o(\kk) \hat G_2(\kk)$ 
and in $d=3$ 
\be \hat G_{i,j;\o}(\kk- \o\bar\z/a)=\d_{i,j,\o}  \chi_\o(\kk) \hat G_{2,i,j} (\kk)
\ee
Similarly we introduce the current correlations
\be
\hat\G_{\m,\m_1,..,\m_n}(\pp_1,..,\pp_n)={\partial^{n+1} 
W \over \partial \hat A_{\m,\pp} \partial \hat A_{\m_1,\pp_1}...\partial \hat A_{\m_n,\pp_n}  }|_0
\quad\hat\G^5_{\m,\m_1,..,\m_n}(\pp_1,..,\pp_n)={\partial^{n+1} 
W \over \partial \hat A^5_{\m,\pp} \partial \hat A_{\m_1,\pp_1}...\partial \hat A_{\m_n,\pp_n}  }|_0
\ee
By Feynman graph expansion one can see that the correlations of \pref{gen}
coincide in the formal limit in which regularizations are removed $a\to 0, M\to 0$ 
with massless QED in the Feynman gauge. The lattice breaks 
the Lorentz symmetry, so that the parameters $t,t'$ have to be chosen as function of 
the coupling $\l$ to
fix the light velocity equal to $c=1$; $\n$ is a counterterm to fix the position of the singularity. The chiral symmetry is also broken and 
one has to 
fix the constants $\ZZ_\m^5$ in order to ensure the following condition, if $\kk=\qq+\o \bar\z/a$, $\qq,\pp$ small, $\o=\pm$
\be
\hat G_{2,1,\m}(\pp,\kk)=\o  \hat G^5_{2,1,\m}(\pp,\kk)(1+O(a \qq,a(\qq+\pp))\label{111}
\ee
The AB non-renormalization means that the anomaly acquires no corrections provided that
the normalizations are fixed so that \pref{111} holds, see e.g. \cite{AB1}.

While the lattice breaks chiral and Lorentz symmetry (which are only emergent), our model respects exactly
gauge symmetry, as by construction
\be
W(A_\m,A^5_\m,\phi)=W(A_\m+\partial_\m \a_\xx,A^5_\m, e^{i\a_\xx}\phi_\xx) \label{xx}
\ee
and from this we get the following Ward Identity  expressing the conservation of the current
\be \pp_\m \hat \G_{\m,\m_1,..,\m_n}=0\label{ss}\ee 
and the relation
\be -i\pp_\m \hat G_{2,1,\m}(\pp,\kk)
=\hat G_2(\kk)-\hat G_2(\kk+\pp)\label{wii}  
\ee
The chiral symmetry is broken by the lattice so that the analogue of 
\pref{ss} for the chiral current is not true. In the emergent continuum theory the chiral symmetry holds exactly but nevertheless 
$\pp_\m \hat \G_{\m,\m_1,..,\m_n}^5$ is non vanishing, what is precisely 
the quantum anomaly \cite{A}. In \cite{NN} it was shown that, in the {\it non-interacting case}, one has in the lattice theory
$\pp_\m \hat\G^5_{\m,\n}(\pp) ={1\over \pi} \e_{\m,\n} \pp_\m$ in $d=1$ and
$\pp_{\mu} \hat\Gamma^5_{\mu,\nu,\sigma}(\pp_{1}, \pp_{2}) =\frac{1}{2\pi^{2}} \pp_{1,\alpha} \pp_{2,\beta} \varepsilon_{\alpha, \beta, \nu, \sigma}$
in $d=3$, that is one gets the same result as the continuum theory. We investigate what happens  to the anomaly in presence of interaction with a finite lattice.
\vskip.3cm
{\bf Theorem.} {\it  For small $\l$ and suitable $\n,t,t'$ and $\ZZ^5_\m$ chosen so that \pref{111} holds,
the correlations of \pref{gen}  are, respectively for $d=1$ and $d=3$ 
\be 
\hat G_\o(\qq)={|\kk|^\h\over Z} g^D_{\o}(\qq)(1+R(\qq))\quad\quad
\hat G_\o(\qq)={1\over Z} g^D_{\o}(\qq)(1+R(\qq))\label{ss2}\ee where
$\h=a\l^2+O(\l^3)$, $Z=1+O(\l)$ and $R(\qq)$ non vanishing and 
$|R(\qq)|\le C a |\qq|$; moreover, up to higher order terms in $\pp$
\be \pp_\m \hat\G^5_{\m,\n}(\pp) ={1\over \pi} \e_{\m,\n} \pp_\m\quad\quad
\pp_{\mu} \hat\Gamma^5_{\mu,\nu,\sigma}(\pp_{1}, \pp_{2}) =\frac{1}{2\pi^{2}} \pp_{1,\alpha} \pp_{2,\beta} \varepsilon_{\alpha, \beta, \nu, \sigma}\label{ss1}
\ee
} 
\vskip.3cm
The above result is an emergent Adler-Bardeen theorem, as \pref{ss1} says that there are no interaction corrections
to the anomaly, even in presence of a finite lattice; 
its value coincides with the one of non interacting Dirac
fermions. In contrast symmetry breaking terms produce non vanishing corrections to the correlations, see \pref{ss2}. The above result is rigorous, as 
the presence of the lattice allows
to get a full non-perturbative control on the functional integrals.

In the rest of the paper a proof of the above result is provided. In \S 3 we describe the Renormalization Group analysis
for the lattice model \pref{gen}, and we get the main regularity properties for the kernels of the effective potential.
In \S 4 we get the anomaly non-renormalization in the $d=3$ case, and in \S 5 in the $d=1$ case; finally \S 6 is devoted to conclusions.

\section{Renormalization Group}

As we are interested in the possible breaking of the AB property due the irrelevant terms, one needs an {\it exact} RG analysis
in order to take them fully into account \cite{fon2},\cite{fon1}.  
The starting point is the decomposition of the propagator in higher and lower energy degrees of freedom, that is
\be 
g(\xx)=g^{(N)}(\xx)+g^{(\le N-1)}(\xx)
\ee
where $\hat g^{(N)}(\kk)$ and $\hat g^{(\le N-1)}(\kk)$ are equal to $\hat g(\kk)$ times $f_N(\kk)$ and $\chi_{N-1}(\kk)$ respectively, where $\chi_{N-1}(\kk)$ is a compact support function
selecting momenta such that $|\kk-\o\bar\z/a|\le \g^N$ with $\g>1$, $\g^N=1/(10 a)$
and $f_N=1-\chi_{N-1}$.
We can use the decomposition property $P(d\psi)=P(d\psi^{(\le N-1)})
P(d\psi^{(N)})$, where $P(d\psi^{(\le N-1)})$ and
$P(d\psi^{(N)})$ have propagator 
$g^{(\le N-1)}(\xx)$ and $g^{(N)}(\xx)$. The field $\psi^{(N)}$ represents the highest energy degree of
freedom; its propagator $g^{(N)}(\xx)$ decays at large distances faster than any power with rate $\g^N$ and is bounded by $\g^{d N}$, 
and it can be integrated out safely. Note that $\chi_{N-1}(\kk)$ as a function of $\kk$
has support in two disconnected regions around $\pm {\bar \z}/a$; we can 
therefore, after shifting the momenta,
write \be g^{(\le N-1)}(\xx)=\sum_{\o=\pm} e^{i \o {\bar \z\over a}\xx}
g_\o^{(\le N-1)}(\xx) \quad\quad \psi^{\pm(\le N-1)}_\xx=
\sum_{\o=\pm} e^{\pm i \o {\bar \z\over a} \xx}\psi^{\pm(\le N-1)}_{\o,\xx}\ee 
In conclusion we get
\bea
&&e^{W(A_\m,A^5_\m,\phi)}=\int P(d\psi^{(\le N-1)})  P(d\psi^{(N)}) 
e^{V^{(N)}(\psi^{(\le N-1)}+\psi^{(N)},\phi,A^5_\m,A_\m)}\nn\\
&&=
\int P(d\psi^{(\le N-1)}) 
e^{V^{(N-1)}(\psi^{(\le N-1)},\phi,A^5_\m,A_\m)}\label{eff1}
\eea
with $V^{(N-1)}(\psi^{(\le N-1)},\phi,A^5_\m,A_\m)$  equal to $\sum_{n=0}^\io  {1\over n!}
E^T_N(V;n)$ and  
$E^T_N$ is the truncated expectation, that is the sum of connected Feynman graphs. 
The effective potential $V^{(N-1)}$ is given by
\be
V^{(N-1)}=\sum_{l,m}\int d\underline\xx  d\underline\yy
W^{(N-1)}_{l,m}(\underline \xx,\underline \yy) [\prod_{i=1}^l \psi^{\e_i (\le N-1)}_{j_i\xx_i,\o_i}][\prod_{i=1}^m A^ {\s_i}_{\m,\yy_i}]
\ee
where $\underline\xx=\xx_1,..,\xx_l$, $\underline\yy=\yy_1,..,\yy_m$, $j=1,2$ in $d=3$ or $j=1$ in $d=1$,
$\e_i=\pm$, $\m=0,1$ in $d=1$ and $\m=0,1,2,3$ in $d=3$, 
$\o=\pm$ and $\s=0,5$ ($A^ {0}_{\m,\yy}\equiv A_{\m,\yy}$).

Note that the RG integration step has two effects; the first is that
the potential is now expressed as sum over monomials of fields of every order
and the second that the field is splitted in two components labeled by $\o=\pm$.
The kernels $W^{(N-1)}_{l,m}$ are expressed by convergent series in $\l$; this follows
from
the representation 
$g^{(N)}(\xx-\yy)=(f_{\xx},g_{\yy})$ where $(,)$ is a suitable scalar product and the fact that
fermionic expectation can be written as the determinant of a Gram matrix $M$ with elements $(f_{\xx_i},g_{\xx_j})$ with bound $|\det M|\le \prod ||f_{\xx_i}||||g_{\xx_i}||$; see e.g.
\cite{GK} or \cite{M1}.
%

We integrate the lower degrees of freedom writing
$g_\o^{(\le N-1)}=\sum_{h=-\io}^{N-1} g^{(h)}_\o$ where 
$g^{(h)}_\o$ has cut-off function $f_h$ with support in $\g^{h-1}\le |\kk\mp {\bar \z}/a |\le \g^{h+1}$;
by integrating the fields $\psi^{(N-1)}_\o,\psi^{(N-2)}_\o,..,\psi^{(h)}_\o$ we get an expression similar to 
\pref{eff1} with  $P(d\psi^{(\le h)})$ with propagator
\be
g^{(\le h)}_{\o}(\xx)=\int {d\kk\over (2\pi)^{d+1}}  {e^{i\kk\xx}\over Z_h}{\chi_h(\kk)\over -i\a_{\m,\o,h} k_\m }+r^h_\o(\xx)\label{lll}
\ee
where $\chi_h=\sum_{k=-\io}^h f_k$ and in $d=1$ one has $\a_{0,\o,h}=1$, $\a_{1,\o,h}=-i \o v_h$
and in
$d=3$ one has $\a_{0,\o,h}= 1,\quad \a_{1,\o,h}=-i\s_1 v_{h,1}  ,\quad \a_{2,\o,h}=-i\s_2 v_{h,1} ,\quad \a_{3,\o,h}=-i\o\s_3  v_{h,3}$; 
the first term is bounded by $\g^{d h}$ and decays faster than any power in 
$\g^h |\xx|$, while the second is smaller, being bounded by $\g^{d h} a \g^{h}$.
The velocities are such that $v_h\to v_{-\io}=v_0+O(\l)$ and we can tune the parameters such that $v_{-\io}=1$. We call $\a_{\m,\o}$ simply $\a_{\m,\o,h}$ with $h=-\io$.
With this choice the first term in the r.h.s. of \pref{lll} is the relativistic propagator at scale $h$.

The effective potential 
$V^h$ can be decomposed in an irrelevant part, containing all the monomials with negative scaling dimension $D=(d+1)-d n/2-m$
, and a relevant and marginal part $D\ge 0$.
The marginal term linear in $A$ have the form 
\be
\sum_{\o=\pm}  \int {d\pp\over (2\pi)^{d+1}}[
Z_{\m,h} \ \hat A_{\m,\pp}\, \hat\jmath_{\m,\o,\pp}+Z_{\m,h}^5 \hat A^5_{\mu,\pp}\, \hat\jmath^5_{\mu,\o,\pp}]\ee
with (in $d=3$ $\hat\psi=(\hat\psi_1,\hat\psi_2)$)
\be
\hat j_{\m,\o,\pp}\!=\!\int \frac{d\kk}{(2\pi)^{d+1}}\hat \psi^+_{\o,\kk+\pp} \a_{\m,\o}\hat\psi^-_{\o,\kk}\quad\quad 
\hat j^{5}_{\m,\o,\pp}\!=\ZZ^5_\m \!\int \frac{d\kk}{(2\pi)^{d+1}}\hat \psi^+_{\o,\kk+\pp} \a_{\m,\o}^{5}\hat\psi^-_{\o,\kk}\quad\quad
\a^5_{\mu,\o}=\o\a_{\mu,\o}\ee 
The factors $Z_{\m,h}$ or 
$Z^5_{\m,h}$ are the renormalizations of the current and axial current respectively.
The relevant term is $\g^h \n_h \sum_\o \int d\xx  \psi^+_{\xx,\o} \bar \a\psi^-_{\xx,\o}$ with $\bar \a=1$ in $d=1$ and $\s_3$ in $d=3$
and $\n$ has to fixed so that so that $\n_h=O(\g^{h-N})$. Finally in $d=1$ there is a marginal interaction 
\be \l_h\int d\xx \psi^+_{\xx,+} \psi^-_{\xx,+}\psi^+_{\xx,-} \psi^-_{\xx,-}\ee which is absent in $d=3$.

The kernels $W^{(h)}_{l,m}$ are obtained, see e.g. \cite{M1}, by contracting the effective potentials at previous scales, and one can distinguish the contributions
$W^{(h)}_{a,l,m}$, obtained contracting 
only marginal terms, from the contributions $W^{(h)}_{b,l,m}$ obtained contracting
at least an irrelevant or relevant $\n$ term; the series expansion are convergent and the following bound holds  \cite{M1}
\be
\int d(\underline\xx/\xx_1) |W^{(h)}_{i,l,m} (\underline \xx)|\le C \g^{D h}  \g^{-\th_i(N-h)} \quad \th_a=0, \th_b=1\quad D=(d+1)-d l/2-m
\label{eff}
\ee
Note that there is an essential difference between the $d=3$ and $d=1$ case; in the first case to 
$W^{(h)}_{a,n,m}$ no vertices with more than two fermionic lines contribute, while in the second also the local vertices quartic in $\psi$ contribute.

The flow of the running coupling constants and renormalizations is quite different. In the $d=3$ case \cite{M2}
the terms with more than 2 fields have negative dimension
so that
\be {Z_{h-1}\over Z_h}=1+O(\l \g^{h-N})\quad v_{h-1}=v_h+O(\l \g^{h-N}) \quad {Z_{\m,h-1}\over Z_{\m,h}}=1+O(\l \g^{h-N})
\quad {Z^5_{\m,h-1}\over Z^5_{\m,h}}=1+O(\l \g^{h-N})\ee
by \pref{eff}. We choose the parameters so that
$v_{h}=1+O(\l \g^{h-N})$.Defining 
$Z_{\m,-\io}\equiv Z_\m$,
$Z^5_{\m,-\io}\equiv Z^5_\m$, $Z_{-\io}\equiv Z$
 we can write 
\be
Z_{\m,h}= Z_\m+O(\l \g^{h-N})\ee and similar expressions for $Z_h, Z_{\m,h}$.

In the $d=1$ case \cite{BFM} in contrast the interaction is marginal and the beta function of the renormalizations is given by
\be {Z_{h-1}\over Z_{h}}=1+a \l_h^2+O(\l_h^3)\ee and similar expressions holds
for $Z_{\m,h}$ and $Z^5_{\m,h}$. It turns out that, as a consequence of the emerging chiral symmetry, the beta function 
for $\l_h$ is asymptotically vanishing $\l_{h-1}=\l_h+O(\l^2 \g^{h-N})$ and the same is true for the velocity.
Note that, as 
$\l_h\to \l_{-\io}=\l+O(\l^2)$, then the renormalization can be singular as $h\to-\io$;
in particular \be Z_h\sim \g^{\h (h-N)}\ee with $\h=-a \l^2+O(\l^3)$. 

The conclusion of the above analysis is that , if we suitable fix the velocities $v_0$ and the counterterms $\n$
one gets \pref{ss2}, that is Lorentz invariance emerges up to corrections which are small if $\qq$ is far from the lattice scale. 


%
%

\section{Anomaly non-renormalization;  the irrelevant case}

In $d=3$ the interaction is irrelevant and, by \pref{eff}, for $\kk\sim \o \bar \z/a$, $\pp\sim 0$, $\o=\pm$ $\hat G_2(\kk)={1\over Z}g(\kk)(1+O(a \qq))$ and
\be
\hat G_{2,1,\m}(\pp,\kk)=Z_\m \hat G_2(\kk) \a_{\m,\o} \hat G_2(\kk+\pp)(1+R)\quad 
\hat G^5_{2,1,\m}(\pp,\kk)=\o \ZZ^5_\m Z^5_\m \hat G_2(\kk) \a_{\m,\o}
\hat G_2(\kk+\pp)(1+R)\ee
with $|R|\le C a (|\qq|,|\qq+\pp|)$. Note the perfect proportionality of the vertex 
function to $Z_\m, Z_\m^5$ which is not true in the marginal case (the $R$ term is not subdominant).
We know from the previous section that  $Z, Z_\m, Z_\m^5$ are expressed by convergent series depending on all details at the lattice scale; 
the Ward Identity \pref{wii}
implies the exact relation \be {Z_\m\over Z}=1\label{112a} \ee A similar identity is not true for $Z_\m^5$ and generically
$Z_\m^5/Z_\m$ is a non trivial function of $\l$. Therefore in order to ensure the validity of 
\pref{111} we choose \be \ZZ^5_\m={Z_\m\over Z^5_\m}\label{112}  \ee
The anomaly coefficient is expressed in terms of
\be\hat\G^5_{\m,\m_1,\m_3}(\pp_1,\pp_2)=\sum_{h=-\io}^N \hat W^{(h)}_{0,3}(\pp_1,\pp_2)
 \ee
By \pref{eff} it is bounded by
\be
|\hat\G^5_{\m,\m_1,\m_3}(\pp_1,\pp_2)|\le C\sum_{h=-\io}^N \g^{h}<\io\ee so that it is {\it continuous} as a function of $\pp_1,\pp_2$; it is however not differentiable as each derivative produces an extra $\g^{-h}$.
The continuity combined with Ward Identites \pref{xx}
are sufficient to prove that $\G^5_{\m,\m_1,\m_2}(0,0)=0$ without any explicit computation: it is sufficient to write from  \pref{xx}
$
\pp_{1,\m_1} \hat \G^5_{\m,\m_1,\m_2}(\pp_1,\pp_2)=0$
at $\pp_{1,1}=\bar p_1$ and zero otherwise  
and use continuity. 
One would be tempted to iterate this argument for the derivative of  $\hat \G^5_{\m,\m_1,\m_2}$, but that is impossibile for the lack of differentiability, and indeed 
$\hat\G^5_{\m,\m_1,\m_2}$ has non vanishing derivatives.

Regularity properties are a very efficient tool to get information on the property of the anomalies, once that  
$\hat\G^5_{\m,\m_1,\m_2}(\pp_1,\pp_2)$ is suitable decomposed in order to get advantage from the dimensional gain in 
\pref{eff}. We write, $\pp=\pp_1+\pp_2$
\be
\hat\Gamma^5_{\mu, \nu, \sigma}(\pp_1,\pp_2)=\media{\hat\jmath^5_{\mu,\pp}; \tilde\jmath_{\n,\pp_1};\tilde\jmath_{\s,\pp_2}}+\D(\pp_1,\pp_2)
\ee
where $\D$ is the Schwinger term and $\tilde j$ the interacting current (obtained by the derivative in $A$). $\D$ has the same bound as the terms with $m=2,1$ hence they are {\it differentiable}. In absence of interaction $\l=0$
$\media{\hat\jmath^5_{\mu,\pp}; \tilde\jmath_{\n,\pp_1};\tilde\jmath_{\s,\pp_2}}$
is expressed by the triangle graph. In presence of interaction, the RG analysis of the previous section says that 
\be
\media{\hat\jmath^5_{\mu,\pp}; \tilde\jmath_{\n,\pp_1};\tilde\jmath_{\s,\pp_2}}=
\sum_{h=-\io}^N \hat W^{(h)}_{a, 0,3}+\sum_{h=-\io}^N \hat W^{(h)}_{b, 0,3}
\ee
where the first term, containing only marginal source terms, is the triangle graph with propagators $g^{(h)}/Z_h$ and vertices associated to $Z_{\m,h},Z^5_{\m,h}$,
while the second is a series of terms with an arbitrary number of quartic interactions, see Fig. 1.
According to the bound \pref{eff} we have \be
\sum_{h=-\io}^N |\partial \hat W^{(h)}_{b, 0,3}|\le  \sum_{h=-\io}^N \g^{ (h-N)}\le C\ee so that 
$\hat W^{(h)}_{b, 0,3}$ is differentiable while $\hat W^{(h)}_{a, 0,3}$ is not.

\insertplot{230}{79}
{\ins{50pt}{40pt}{$=$}
\ins{130pt}{40pt}{$+$}
\ins{210pt}{40pt}{$+...$}
}
{figjsp44a}
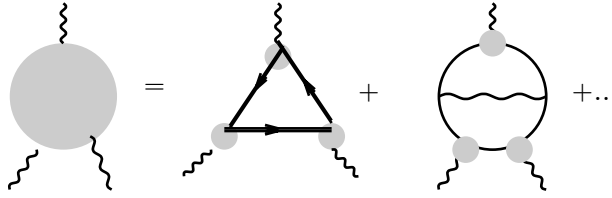
{\label{h2} The decomposition of
$\media{ \hat\jmath^5_{\mu,\pp}; \tilde\jmath_{\n,\pp_1};\tilde\jmath_{\s,\pp_2}}$.
} {0}
We can replace in the renormalized triangle graph the values of $Z_{\mu,h},Z^5_{\mu,h}, v_h$ with their limiting value;
the difference has again an extra $O(\g^{h-N})$ so gives a differentiable contribution. Summing over the scale $h$ has the effect that the cut-off $f_h$ of
single scale propagators add up to $\chi=\sum_{h=-\io}^N f_h$
so that we get at the end
\be
\sum_{h=-\io}^N W^{(h)}_{a, 0,3}=\frac{\ZZ_\m^5 Z^5_{\m} Z_{\n} Z_{\s}}{Z^3 }\, I_{\m,\n,\s}(\pp_1,\pp_2)+G(\pp_1,\pp_2)\ee
where the second term is differentiable while $I_{\m,\n,\s}(\pp_1,\pp_2)$ is the relativistic
triangle graph with propagators ${\chi(\kk)\over -i \not\kk} $, that is with a momentum cut-off.
In conclusion
\be
\hat\G^5_{\mu,\n,\s}(\pp_1,\pp_2)=
\frac{\ZZ_\m^5 Z^5_{\m} Z_{\n} Z_{\s}}{Z^3 }\, I_{\m,\n,\s}(\pp_1,\pp_2)
+H^5_{\mu,\n,\s}(\pp_1,\pp_2)\label{dec}\ee
{where} $H^5_{\m,\n,\s}$ is continuously differentiable.   By \pref{112a},\pref{112} we get 
\be
\frac{\ZZ_\m^5 Z^5_{\m} Z_{\n} Z_{\s}}{Z^3 }=1\ee 
In addition the contribution from the first term in \pref{dec} can be explicitly computed, see
\cite{GMP}, and one gets 
\be
\pp_\mu I_{\m,\nu,\sigma}(\pp_{1}, \pp_{2}) = \frac1{6\pi^2} \pp_{1,\alpha} \pp_{2,\beta} 
\varepsilon_{\alpha \beta \nu\sigma }\quad\quad 
\pp_{1,\nu} I_{\m,\nu,\sigma}( \pp_{1},\pp_{2})=\frac1{6\pi^2} \pp_{1,\alpha} \pp_{2,\beta} \varepsilon_{\alpha\beta \mu \sigma}\label{21} \ee
up to higher order terms, cubic in the momenta; moreover
$
\pp_{2,\sigma}I_{\mu,\nu,\sigma}(\pp_{1}, \pp_{2}) = \pp_{2,\sigma} I_{\mu,\sigma,\nu}(\pp_{2}, \pp_{1}) = \frac1{6\pi^2}
\pp_{2,\alpha} \pp_{1,\beta} \varepsilon_{\alpha, \beta, \mu, \nu}$.  Note that the r.h.s. 
of \pref{21}
do not depend on the cut-off $1/a$; moreover
either the current and the chiral current are not conserved in $I_{\mu,\nu,\sigma}$ as the momentum cut-off breaks the local gauge invariance.

It remains to evaluate the second term in \pref{dec}; it depends on all the irrelevant terms and is expressed by a complicate series so it cannot be explicitly computed;
however we show now that the information that is differentiable combined with Ward Identity \pref{xx} is sufficient for its determination.
Indeed from the WI \pref{xx} we get   
\be \pp_{1,\nu} \hat\G^5_{\m,\nu,\sigma}(\pp_{1}, \pp_{2})=0\ee 
We use now the decomposition \pref{dec} and the differentiability of 
$H^5_{\mu,\nu,\sigma}$ to expand up to first order 
\begin{equation}
0 =  \frac1{6\pi^2} \pp_{1,\alpha} \pp_{2,\beta} \varepsilon_{\alpha,\beta,\mu,\sigma} + \pp_{1,\nu} \pp_{1,\alpha} \frac{\partial}{\partial \pp_{1,\alpha}} H^5_{\mu,\nu,\sigma}(\V0,\V0) + \pp_{1,\nu} \pp_{2,\beta} \frac{\partial}{\partial \pp_{2,\beta}} H^5_{\mu,\nu,\sigma}(\V0,\V0) + O(\pp^{3})
\end{equation}
From the above relation we get
$\frac{\partial}{\partial\pp_{1,\alpha}} H^5_{\mu,\nu,\sigma}(\V0,\V0) + \frac{\partial}{\partial\pp_{1,\nu}} H_{\mu,\alpha,\sigma}(\V0,\V0)= 0$ and
\be
\frac1{6\pi^2}  \varepsilon_{\alpha, \beta, \mu,\sigma} = -\frac{\partial}{\partial \pp_{2,\beta}} H^5_{\mu,\alpha,\sigma}(\V0,\V0)
\ee
Similarly from $
\pp_{2,\sigma} \hat\Gamma_{\mu,\nu,\sigma}(\pp_{1}, \pp_{2}) = 0$ we get
\begin{equation}
0 =\frac1{6\pi^2} \pp_{2,\alpha} \pp_{1,\beta} \varepsilon_{\alpha, \beta, \mu, \nu} + \pp_{2,\sigma} \pp_{1,\alpha} \frac{\partial}{\partial \pp_{1,\alpha}} H_{\mu,\nu,\sigma}(\V0,\V0) + \pp_{2,\sigma} \pp_{2,\beta} \frac{\partial}{\partial\pp_{2,\beta}} H_{\mu,\nu,\sigma}(\V0,\V0) + O(\pp^{3})\;,
\end{equation}
and $\frac1{6\pi^2}\varepsilon_{\alpha, \beta, \mu, \nu} = -\frac{\partial}{\partial\pp_{1,\beta}} H_{\mu,\nu,\alpha}(\V0,\V0)$.
Finally
\be
H_{\mu,\nu,\sigma}(\pp_{1}, \pp_{2}) = \pp_{1,\alpha} \frac{\partial}{\partial \pp_{1,\alpha}} H_{\mu,\nu,\sigma}(\V0, \V0) + \pp_{2,\beta} \frac{\partial}{\partial \pp_{2,\beta}} H_{\mu,\nu,\sigma}(\V0, \V0)
= -\frac1{6\pi^2}  \pp_{1,\alpha} \varepsilon_{\sigma, \alpha, \mu,\nu} -\frac1{6\pi^2}  \pp_{2,\beta} \varepsilon_{\nu, \beta, \mu, \sigma}\ee
so that
\be \pp_{\mu} \hat\Gamma^5_{\mu,\nu,\sigma}(\pp_{1}, \pp_{2}) =\frac1{6\pi^2}(
\pp_{1,\alpha} \pp_{2,\beta} \varepsilon_{\alpha,\beta, \nu, \sigma}  -\pp_{1,\alpha}\pp_{2,\mu} \varepsilon_{\sigma, \alpha, \mu,\nu}-\pp_{2,\beta}\pp_{1,\mu} \varepsilon_{\nu,\beta, \mu,\sigma}) = \frac{1}{2\pi^{2}} \pp_{1,\alpha} \pp_{2,\beta} \varepsilon_{\alpha, \beta, \nu, \sigma}
\ee
up to higher orders terms in $\pp$. This says that the AB non-renormalization property holds even in presence of symmetry breaking terms.

\section{Anomaly non-renormalization; marginal interactions}

We have derived in the previous section the AB non-renormalization in a case where the interaction is irrelevant;
this is in contrast with the $d=3$ case with massless photons where the interaction is marginal. However we show now that even in $d=1$,
where the interaction is marginal, the AB renormalization holds exactly.
Again we can decompose 
\be
\hat\G^5_{\mu,\n}(\pp)=
\hat\G^{5,a}_{\mu,\n}(\pp)+\hat\G^{5,b}_{\mu,\n}(\pp )\label{sao} \ee
where $\hat\G^{5,a}_{\mu,\n}$ contains only marginal 
terms and $\hat\G^{5,b}_{\mu,\n}$ at least an irrelevant or relevant term; therefore by \pref{eff}
we get
\be
|\hat\G^{5,i}_{\mu,\n}(\pp)|\le \sum_{h=0}^N \g^{\th_i(h-N)}
\ee
with $\th_a=0, \th_b=1$; therefore
the first term in \pref{sao}
is not continuous while the first is continuous. 
$\hat\G^{5,a}_{\mu,\n}(\pp)$ has a relativistic structure and we could try to follow 
a strategy similar to the one in the previous section.
There is however a major difference; now there are marginal terms
quartic in the fields, so that the first term is expressed as a series of renormalized graphs and not by a single term. As the dominant part now corresponds to an interacting theory,
it seems that it cannot be explicitly computed. We can however introduce a relativistic QFT 
describing Dirac fermions in $d=1+1$ with a current-current non local interaction; the corresponding generating function is given by
\be
e^{W_{rel}(A,A^5,\phi)}  =\int P(d\psi^{\le K}) e^{\l_\io  \tilde Z^2 \int d\xx d\yy  v(\xx-\yy) j_{\m,\xx} j_{m,\yy}+\sum_\m \tilde Z_\m 
\int d\xx A_\m j_\m+\sum_\m \tilde Z^5_\m \int d\xx A^5_\m j^5_\m+\int d\xx (\psi^+_\xx\phi^-_\xx+\psi^-_\xx\psi^+_\xx))
}\label{em}
\ee
where $\psi^\pm_{\xx,\o}$, $\o=\pm$ are Grassmann variables, $j_{0,\xx}=\psi^+_{\xx,+}\psi^-_{\xx,+}+\psi^+_{\xx,-}\psi^-_{\xx,-}$, $j_{1,\xx}=
i(\psi^+_{\xx,+}\psi^-_{\xx,+}-\psi^+_{\xx,-}\psi^-_{\xx,-})$,
$P(d\psi^{\le K}) $ has propagator, if $\o=\pm$,
$\hat g_\o^{(\le K)}(\kk)=
{\chi_K(\kk)\over \tilde Z (-i k_0+\o k)}$ with $\chi_K(\kk)$ a cut-off function non vanishing for $|\kk|\le \g^K$
and $v(\xx-\yy)$ decaying exponentially with rate $1/a$. This theory is in a sense the regularization of the scaling limit of the previous one, and it verifies
the chiral global gauge invariance (which is broken by the lattice).  

The RG analysis of \pref{em} is similar to the one in \S 3 and we can choose
the parameters
$\tilde Z, \tilde Z^5_\m, \tilde Z_\m, \l_\io$ in \pref{em} as function of $\l$ so that the limiting value at $h=-\io$ of the corresponding running coupling constants is the same as in the lattice theory.
By this choice the difference in the running coupling constants is $O(\g^{h-N})$ so that
we get the decomposition
\be
\hat\G^5_{\mu,\n}(\pp)=\ZZ^5_\m [{\partial^2 W_{rel}(A,A^5,\phi)\over\partial \hat A^5_\m \partial \hat A_\n}|_0
+H^5_{\mu,\n}(\pp)]\quad \hat\G_{\mu,\n}(\pp)= [{\partial^2 W_{rel}(A,A^5,\phi)\over\partial \hat A_\m \partial \hat A_\n}|_0
+H_{\mu,\n}(\pp)]\label{dec1}
\ee
where
$H_{\mu,\n}(\pp),H^5_{\mu,\n}(\pp)$ continuous by \pref{eff}; similarly, up to subdominant terms in the momentum,
\be \hat G_{2,\mu}={\partial^2 W_{rel}\over\partial \hat\phi^+\partial\hat\phi^-}\quad\quad
\hat G_{2,1,\mu}={\partial^3 W_{rel}\over\partial \hat A_\m \partial \hat\phi^+\partial\hat\phi^-}\quad\quad
\hat G^5_{2,1,\mu}=\ZZ_\m^5{\partial^3 W_{rel}\over\partial \hat A^5_\m \partial \hat\phi^+\partial\hat\phi^-}
\ee

We can take advantage from the fact that the 
model \pref{em} verifies global and axial symmetries; however local symmetries are broken by the presence of the momentum cut-off and this produces
extra anomalous terms in the WI for the global and axial current.
Note indeed that, if $D_\o(\kk)=-i k_0+\o k$ 
\be
\hat g_\o^{(\le K)}(\kk)-\hat g^{(\le K)}_\o(\kk+\pp)-\hat g^{(\le K)}_\o(\kk) D_\o(\pp) g^{(\le K)}_\o(\kk+\pp)=\hat g^{(\le K)}_\o(\kk)C(\kk,\pp) \hat g^{(\le K)}_\o(\kk+\pp)
\ee
with $C(\kk,\pp) =D_\o(\kk) (\chi_K^{-1}(\kk)-1)-D_\o(\kk+\pp)(\chi_K^{-1}(\kk+\pp)-1)$ (the r.h.s.would be zero in absence of cut-off).
The presence of this extra term produce an additional factor in the WI, see Fig. 2; as proven in 
\cite{BFM} in the $K\to\io$ limit the following WI for the vertex and chiral vertex are obtained
\bea
&&-i p_0  {1\over \tilde Z_0}{\partial^3 W_{rel}\over\partial \hat A_{0,\pp}\partial \hat\phi^-_{\kk,\o}\partial\hat\phi^+_{\kk+\pp,\o}}
+{p_1\over \tilde Z_1} {\partial^3 W_{rel}\over\partial \hat A_{1,\pp}\partial \hat \phi^-_{\kk,\o}\partial\hat\phi^+_{\kk+\pp,\o}}={1\over \tilde Z(1-\t)} 
({\partial^2 W_{rel}\over\partial \hat\phi^-_{\kk,\o}\partial\hat\phi^+_{\kk,\o} }-{\partial^2 W_{rel}\over\partial \hat\phi^-_{\kk+\pp,\o}\partial\hat\phi^+_{\kk+\pp,\o}})
\label{sss}\\
&&-i p_0  {1\over \tilde Z_0^5 }{\partial^3 W_{rel}\over\partial \hat A^5_{0,\pp}\partial \phi^-_{\kk,\o}\partial\phi^+_{\kk+\pp,\o}}
+{ 
p_1\over  \tilde Z^5_1}{\partial^3 W_{rel} \over\partial \hat A^5_{1,\pp}\partial \phi^-_{\kk,\o}\partial\phi^+_{\kk+\pp,\o}}
={\o
\over \tilde Z(1+\t)}
( {\partial^2 W_{rel}\over\partial \hat\phi^-_{\kk,\o}\partial\hat\phi^+_{\kk,\o}}-{\partial^2 W_{rel}\over\partial \hat\phi^-_{\kk+\pp,\o}\partial\hat\phi^+_{\k+\pp,\o
}})\nn   
\eea
and $\t=\l_\io/4\pi$. The extra term in the WI produced by the $C$- term reduces, in the limit $K\to\io$, to the vertex function times the constant $\t$ (which is the graph 
for the anomaly in $d=1$ with momentum cut-off).

The fact that the vertex and 2-point function of \pref{sss} and lattice model (computed at $\qq+\o \z/a$ with $\qq$ small)
are close up to $O(a \qq)$ terms says that the first of the WI 
\pref{sss} coincides with \pref{wii}; this imposes constraints for the parameters of effective QFT 
\pref{em}, that is  
\be
{\tilde Z_1\over \tilde Z_0}=1\quad \quad {\tilde Z_0\over \tilde Z}=1-\t\label{conf}
\ee
\insertplot{280}{79}
{\ins{52pt}{40pt}{$=$}\ins{124pt}{40pt}{$-$}\ins{190pt}{40pt}{$+$}
}
{figjsp44b2}
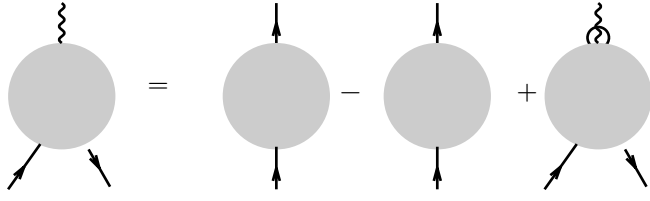
{\label{h2} The WI for the vertex function of \pref{em} where the last term is the extra term due to the $C$ factor.
} {0}
%
%
%
%
%
We have now to choose $\ZZ_\m^5$ by \pref{111}; from 
\pref{sss} in the limit $p_0\to 0, p\to0$
\be
\hat G^5_{2,1}=i \o {\ZZ_0^5 \tilde Z^5_0 \over \tilde Z(1+\t)}\partial_0 {\partial^2 W_{rel}\over\partial \hat\phi^-_{\kk,\o}\partial\hat\phi^+_{\kk,\o} }
\quad 
\hat G_{2,1}=i {\tilde Z_0 \over \tilde Z(1-\t)}\partial_0 {\partial^2 W_{rel}\over\partial \hat\phi^-_{\kk,\o}\partial\hat\phi^+_{\kk,\o} }
\ee
and a similar expression for $\m=1$ so that
%
%
\be
\ZZ^5_i={1+\t\over 1-\t} {\tilde Z_i\over \tilde Z_i^5}=(1+\t){\tilde Z\over \tilde Z_i^5}
\label{29}
\ee
The WI for the current correlations of \pref{em} are
\be
\sum_\m \pp_\m {\tilde Z\over \tilde Z^5_\m}{\tilde Z\over \tilde Z_\n}{\partial^2 W_{rel}\over\partial \hat A_\m^5 \partial \hat A_\n}={\e_{\m,\n}\pp_\m\over 1+\t}{1\over 2\pi}\quad\quad
\sum_\n \pp_\n {\tilde Z\over \tilde Z^5_\m}{Z\over Z_\n}{\partial^2 W_{rel}\over\partial \hat A^5_\m \partial \hat A_\n}={\e_{\n,\m}\pp_\n\over 1-\t}{1\over 2\pi }
\ee
and from \pref{conf}, \pref{29}
\be
{1\over 1-\t}\sum_\m \pp_\m \ZZ^5_\m 
{\partial^2 W_{rel}\over\partial \hat A_\m^5 \partial \hat A_\n}=\e_{\m,\n}\pp_\m {1\over 2\pi }\quad\quad
{1\over 1+\t} \sum_\n \pp_\n \ZZ^5_\m{\partial^2 W\over\partial \hat A^5_\m \partial \hat A_\n}=\e_{\r,\m}\pp_\r {1\over 2\pi}
\ee
Now we use that the lattice Ward identity \pref{xx} and the decomposition \pref{dec1}
\be
\pp_\n \hat\G^5_{\m,\n}=   \ZZ^5_\m \sum_\n \pp_\n  [{\partial^2 W\over\partial \hat A^5_\m \partial \hat A_\n}+H_{\m,\n}]=0
\ee
from which we get
\be
\e_{\r,\m}{(1+\t)\over \ZZ^5_\m } 
\pp_\r {1\over 2\pi }+\pp_\n  H_{\m,\n}(\pp)=0
\ee
In contrast with $\G^5_{\m,\n}(\pp)$, we know that 
$H_{\m,\n}(\pp)$ is continuous in $\pp$ so that  
\be
-\e_{\r,\m}{(1+\t)\over \ZZ^5_\m} 
{1\over 2\pi }=H_{\m,\r}(0)
\ee
and, up to higher orders in $\pp$
\bea
&&\pp_\m \hat\G^5_{\m,\n}(\pp) =\sum_\m
\pp_\m \ZZ_\m^5
[{\partial^2 W_{rel}\over\partial \hat A^5_\m \partial \hat A_\n}+H_{\m,\n}]
=\nn\\
&&
\e_{\m,\n}{(1-\t)\over 2\pi }\pp_\m 
-\e_{\n,\m}{(1+\t)\over 2\pi } \pp_\m =
[(1-\t)+(1+\t)]\e_{\m,\n} \pp_\m {1\over 2\pi }=\e_{\m,\n} \pp_\m {1\over \pi}
\eea
so that the factor $\t$, depending on $\l$, cancels out 
and also in the marginal case the anomaly is non-renormalized.

\section{Conclusions}

The renormalizability of the Standard Model relies on the 
AB non renormalization property which is used in the anomaly cancellation.
It is therefore interesting to see if the anomaly non-renormalization holds generically even
when symmetry breaking terms are present at the Planck scale, or if in contrast its validity requires that they are absent or at least of special form.
We have investigated such a question in 
QED lattice model 
both when the interaction is irrelevant or marginal, showing that 
the AB property holds exactly even if 
Lorentz or chiral symmetry is broken and corrections to correlations are present. The fact that the corrections to the anomaly are due to irrelevant terms requires the use of exact and non-perturbative RG methods.
It would be interesting to establish a similar property removing the mass regularization of
photons; in such a case fermionic cancellations are not sufficient to achieve convergence and 
large/small field decomposition is necessary to get non-perturbative results.

\medskip

{\bf Acknowledgements.} This work has been supported by MIUR, PRIN 
2017 project MaQuMA. PRIN201719VMAST01.

\end{document}